\documentclass[aps,prd,twocolumn,floatfix,preprintnumbers,nofootinbib,superscriptaddress]{revtex4}

\usepackage{amsmath}
\usepackage{graphicx}
\usepackage{textcomp}
\usepackage{amsfonts}
\usepackage{amssymb}
\usepackage{bm}
\usepackage[dvips]{color}

\usepackage{booktabs}

\begin{document}

\title{The $\Omega_{cc}$ resonances with negative parity in the chiral constituent quark model}

\author{Jin-Bao Wang}
\affiliation{School of Physical Science and Technology, Southwest
University, Chongqing 400715, China}

\author{Gang Li}\email{gli@qfnu.edu.cn}
\affiliation{College of Physics and Engineering, Qufu Normal
University, Qufu 273165, China}

\author{Cheng-Rong Deng}\email{crdeng@swu.edu.cn}
\affiliation{School of Physical Science and Technology, Southwest
University, Chongqing 400715, China}

\author{Chun-Sheng An}\email{ancs@swu.edu.cn}
\affiliation{School of Physical Science and Technology, Southwest
University, Chongqing 400715, China}

\author{Ju-Jun Xie}\email{xiejujun@impcas.ac.cn}
\affiliation{Institute of Modern Physics, Chinese Academy of
Sciences, Lanzhou 730000, China} \affiliation{School of Nuclear
Science and Technology, University of Chinese Academy of Sciences,
Beijing 101408, China} \affiliation{School of Physics and
Microelectronics, Zhengzhou University, Zhengzhou, Henan 450001,
China} \affiliation{Lanzhou Center for Theoretical Physics, Key
Laboratory of Theoretical Physics of Gansu Province, Lanzhou
University, Lanzhou 730000, China}

\date{\today}

\begin{abstract}

Spectrum of the low-lying $\Omega_{cc}$ resonances with negative
parity, which are assumed to be dominated by $sccq\bar{q}$
pentaquark components, is investigated using the chiral constituent
quark model. Energies of the $\Omega_{cc}$ resonances are obtained
by considering the hyperfine interaction between quarks by
exchanging Goldstone boson. Possible $sccq\bar{q}$ configurations
with spin-parity $1/2^{-}$, $3/2^{-}$ and $5/2^{-}$ are taken into
account. Numerical results show that the lowest $\Omega_{cc}$
resonances with negative parity may lie at $4050 \pm 100$~MeV. In
addition, the transitions of the $\Omega_{cc}$ resonance to a
pseudoscalar meson and a ground baryon state are also investigated
within the chiral Lagrangian approach. We expect that these
$\Omega_{cc}$ resonances could be observed in the $D\Xi_{c}$
channel by future experiments.

\end{abstract}

\maketitle

\section{Introduction}
\label{intro}
%
%
Recently, five narrow $\Omega_{c}^{0}$ resonances and a doubly
charmed baryon $\Xi_{cc}^{++}$ were observed by LHCb
collaboration~\cite{Aaij:2017nav,Aaij:2017ueg,Aaij:2018wzf}. These
observations are of significant importance in hadron physics since
the experimental data forbaryon resonances with one or more charm
quarks are very poor before 2017~\cite{Tanabashi:2018oca}.
Accordingly, the theorists also made great efforts to describe the
spectrum and decay behaviours of the observed heavy baryon
resonances with charm quarks using various of approaches, such as
the quark model within the three- or five-quark
picture~\cite{Karliner:2017kfm,Wang:2017hej,Wang:2017vnc,Chen:2017gnu,Lu:2017meb,Xiao:2017udy,Yang:2017rpg,Huang:2017dwn,An:2017lwg},
QCD sum
rules~\cite{Chen:2017sci,Wang:2017zjw,Wang:2017xam,Chen:2017sbg,Mao:2017wbz},
the chiral perturbation
approach~\cite{Xiabng:2018qsd,Guo:2017vcf,Yao:2018ifh,Li:2017pxa},
the lattice
QCD~\cite{Padmanath:2017lng,Bahtiyar:2018vub,Mathur:2018rwu}, etc.
Furthermore, the molecular nature of these heavy baryon resonances
were also studied in
Refs.~\cite{Debastiani:2017ewu,Liang:2017ejq,Yu:2019yfr,Dias:2019klk},
where they were dynamically generated from the meson-baryon
interactions in the coupled channels. Besides, it is also very
interesting that the observation of doubly heavy baryon is claimed
to imply existence of heavy tetraquark
mesons~\cite{Karliner:2017qjm,Eichten:2017ffp}. Thus, it is
appropriate to perform further studies about these heavy baryons.

On the other hand, three hidden-charm $N_{c\bar{c}}$ pentaquark
states were observed by LHCb collaboration in recent
years~\cite{Aaij:2015tga,Aaij:2019vzc}, where the experimental data
are in agreement with the predictions made in
Refs.~\cite{Wu:2010jy,Wu:2010vk}. Consequently, rather than the
three-quark picture, it may be more interesting to study the
properties of baryon resonances near or above $4$~GeV within a
five-quark picture, since the energy for pulling a light
quark-antiquark pair to form a pentaquark configuration as the
baryon excitation may be lower than that for the traditional orbital
and radial excitations of a three-quark
configuration~\cite{An:2008xk}. Taking the five-quark picture, the
$\Omega$ excited states with negative
parity~\cite{Yuan:2012zs,An:2013zoa,An:2014lga,Xu:2015bpl,Lin:2018nqd,Huang:2018wth,Pavao:2018xub,Liu:2020yen},
nucleon excited
states~\cite{Gao:2017hya,Huang:2018ehi,He:2018plt,Li:2017kzj,An:2018vmk,An:2009uv,Lin:2018kcc},
and the newly observed $\Omega_{c}^{0}$
resonances~\cite{Yang:2017rpg,Huang:2017dwn,An:2017lwg} are
investigated explicitly. Suggestions on how to observe the
$\Omega(2012)$ state by looking at $\Omega_c$ weak decay process
have been made in Ref.~\cite{Zeng:2020och}. It was found that the
observed small energy splitting of the $\Omega_{c}^{0}$
resonances~\cite{An:2017lwg}, the masses and decay behaviours of the
observed $\Omega(2012)$~\cite{Lin:2018nqd,Huang:2018wth,Pavao:2018xub,Liu:2020yen}
can be well
described by taking either the hadronic molecule picture or the
compact pentaquark configuration,
while it's of course that one cannot
rule out the three-quark components in the baryon
resonances.

The observation of the $\Xi_{cc}$ states have brought new
opportunities for us to study the doubly charmed baryons, since this
finding suggests the potential of discovering more low-lying doubly
charmed baryons in the near future, and thus one needs to have a
solid theoretical calculations for the corresponding spectrum. In
the present work, based on the chiral constituent quark model, we
study the spectrum of the low-lying $\Omega_{cc}$ resonances with
negative parity. And the transitions of these
$\Omega_{cc}$ states to a pseudoscalar meson and a ground baryon
state ($MB$) are also studied, employing the chiral
Lagrangian approach which has been explicitly developed to study the
strong decays of the $N_{s\bar{s}}$ nucleon resonances as in
Ref.~\cite{An:2018vmk}.

The present manuscript is organized as following: in
section~\ref{theo}, we briefly present our theoretical formalism
which includes the Hamiltonian and wave functions for the
$\Omega_{cc}$ pentaquark system, and the chiral Lagrangian approach
for strong decays of a five-quark system, we show our explicit
numerical results in section~\ref{num}, and section~\ref{sumcon}
contains summary and conclusions.

%
%
\section{Theoretical Frame}
\label{theo}

We will briefly introduce the Hamiltonian and wave
functions for the $\Omega_{cc}$ resonances with negative parity as
pentaquark states in Sec.~\ref{ham}, and the chiral Lagrangian
approach for strong decays of the $\Omega_{cc}$ states in
Sec.~\ref{cla}.

\subsection{Hamiltonian and wave functions}
\label{ham}

In present work, the constituent quark model is employed to study
the spectrum of $\Omega_{cc}$ resonances, within which the
Hamiltonian for a five-quark system can be written as

\begin{equation}
H=\sum_{i<j}^{5}H_{hyp}^{ij}+\sum_{i=1,5}m_{i}+H_{o}\,,
\label{ham5}
\end{equation}
where $H_{hyp}^{ij}$ represents the hyperfine interaction between
the $i$th and $j$th quarks in the five-quark system, $m_{i}$ is the
constituent mass of the $i$th quark, and $H_{o}$ is the Hamiltonian
concerning orbital motions of the quarks, which should contain the
kinetic term, the confinement potential of the quarks, and the
flavor symmetry breaking term.

In general, the corresponding
eigenvalue $E_{0}$ of $H_{o}+\sum_{i=1,5}m_{i}$ in Eq.~(\ref{ham5})
should depend on the constituent masses of quarks
and the model parameters in the quark confinement model, for
instance, the confinement strength $C$ and constant $V_{0}$ in the
harmonic oscillator potential model~\cite{Glozman:1995fu}. In
this work, we study the low-lying $\Omega_{cc}$ resonances with
negative parity as pentaquark states, which require all the quarks
and antiquark to be in their ground states, accordingly, the
eigenvalue $E_{0}$ should be the same one for different pentaquark
configurations.

The parameter $E_{0}$ has been taken to be
$2127$~MeV for investigations on the intrinsic sea flavor content of
nucleon in Ref.~\cite{An:2012kj}, with which value the data for light sea
quark asymmetry $\bar{d}-\bar{u}$ in the proton can be well reproduced,
while to fit the experimental data about $\Omega_{c}^{0}$
resonances, we took $E_{0}=3132$~MeV in Ref~\cite{An:2017lwg}.
As discussed in details in Ref.~\cite{An:2017lwg},
the resulted different values of $E_{0}$ by fitting
the experimental data should be consistent with
the chiral constituent quark model if all the model parameters
are taken to be the empirical values.

In this work, the value of $E_{0}$
should be $\sim 1140$~MeV higher than the one we took in
Ref.~\cite{An:2017lwg}, because of the different quark content
in $\Omega_{cc}$ and $\Omega_{c}$ sates, while the $SU(4)$ flavor symmetry breaking
effects caused by two charm quarks in present case will lower
$E_{0}$ by $\sim 170$~MeV than those caused by one charm quark as in
Ref.~\cite{An:2017lwg}, if the Hamiltonian for symmetry breaking
correction is taken to be the form similar as in
Ref.~\cite{An:2013zoa}. Consequently, hereafter we will take $E_{0} =
4102~\mathrm{MeV}$, based on the investigations on the intrinsic
sea content of nucleon and spectrum of low-lying $\Omega_{c}^{0}$
resonances, and the requirements of the chiral constituent quark model.
Nevertheless, we will investigate the dependency of the results
on $E_{0}$.

The hyperfine interaction between quarks is taken to be mediated by
goldstone boson exchange and the corresponding $H_{hyp}^{ij}$ is
taken as following
\begin{eqnarray}
H^{ij}_{hyp}&=&-\vec{\sigma}_{i}\cdot\vec{\sigma}_{j}
                    \Big[ \sum_{a=1}^{3}V_{\pi}(r_{ij})\lambda^{a}_{i}\lambda^{a}_{j}+
                   \sum_{a=4}^{7}V_{K}(r_{ij})\lambda^{a}_{i}\lambda^{a}_{j}\nonumber\\
                  &&+V_{\eta}(r_{ij})\lambda^{8}_{i}\lambda^{8}_{j}
                  +\sum_{a=9}^{12}V_{D}(r_{ij})\lambda^{a}_{i}\lambda^{a}_{j}\nonumber\\
                 &&+\sum_{a=13}^{14}V_{D_{s}}(r_{ij})\lambda^{a}_{i}\lambda^{a}_{j}
                  +V_{\eta_{c}}(r_{ij})\lambda^{15}_{i}\lambda^{15}_{j}\Big] \,,
\label{hyp}
\end{eqnarray}
where $V_{M}(r_{ij})$ denotes the coupling strength for a meson $M$
exchanged between the $i$th and $j$th quarks. In this work, the
$\pi$, $K$, $\eta$, $D$, $D_{s}$ and $\eta_{c}$ mesons are
taken into account.

For a five-quark system with the quark flavor as $\Omega_{cc}$
resonances, namely, the $sccq\bar{q}$ system, a general wave
function can be written as
\begin{eqnarray}
\psi_{t,J_z}^{i} &=&\sum_{a,b,c}\sum_{Y,y,T_z,t_z}\sum_{S_z,s_z}
C^{[1^4]}_{[31]_a[211]_{\bar{a}}} C^{[31]_a}_{[F^{i}]_b [S^{(i)}]_c}[F^{i}]_{b,Y,T_z}\nonumber\\
&& [S^{i}]_{c,S_z}
[211;C]_a\langle Y,T,T_z,y,\bar t,t_z|0,0,0\rangle\nonumber\\
&&\langle S,S_z,1/2,s_z|J,J_z\rangle\bar\chi_{y,t_z}\bar\xi_{s_z}\varphi(\{\vec{\xi}_{j}\})\, .
\label{wfc}
\end{eqnarray}
where $[F^{i}]_{b,Y,T_z}$, $[S^{i}]_{c,S_z}$ and $[211;C]_a$ are the
flavor, spin and color wave functions of the four-quark subsystem
denoted by Young tableaux, the label $i$ enumerates different
pentaquark configurations. The $\vec{\xi}_{j}$ is the Jacobi
coordinates for a five quark system, which is defined as
\begin{eqnarray}
\vec{\xi}_j&=&\frac{1}{\sqrt{j+j^{2}}}\Big(\sum_{i=1}^{j}\vec{r}_{i}-j\vec{r}_{j+1}\Big),
j=1,\cdots,4\, .\label{Jacobi}
\end{eqnarray}

According to the $SU(2)$ symmetry, the spin wave function of a
four-quark system may be $[4]_{S}$, $[31]_{S}$ or $[22]_{S}$, the
corresponding spin quantum numbers are $2$, $1$ and $0$,
respectively. While the coupling between spin of the four-quark
subsystem and the antiquark leads to $J=1/2$, $3/2$ or $5/2$. Given
that all the quarks and antiquark are in their ground states, namely, the orbital
wave function of the four-quark system is $[4]_{X}$, then the flavor wave
function of the $sccq$ subsystem can be $[4]_{F}$, $[31]^1_{F}$,
$[31]^2_{F}$, $[22]_{F}$ and $[211]_{F}$. Finally, the possible
pentaquark configurations denoted by $|i\rangle$ for spin-parity
quantum number $J^P=1/2^{-}$ are
\begin{eqnarray}
|\hspace{0.09cm}1\hspace{0.09cm}\rangle:&sccq_{[4]_{X}[211]_{F}[22]_{S}[211]_{C}}\bar{q}\,,\nonumber\\
|\hspace{0.09cm}2\hspace{0.09cm}\rangle:&sccq_{[4]_{X}[31]^1_{F}[22]_{S}[211]_{C}}\bar{q}\,,\label{1half}\\
|\hspace{0.09cm}3\hspace{0.09cm}\rangle:&sccq_{[4]_{X}[31]^2_{F}[22]_{S}[211]_{C}}\bar{q}\,,\nonumber
\end{eqnarray}
for $J^P=1/2^{-}$ or $3/2^{-}$ are
\begin{eqnarray}
|\hspace{0.09cm}4\hspace{0.09cm}\rangle:&sccq_{[4]_{X}[211]_{F}[31]_{S}[211]_{C}}\bar{q}\,,\nonumber\\
|\hspace{0.09cm}5\hspace{0.09cm}\rangle:&sccq_{[4]_{X}[22]_{F}[31]_{S}[211]_{C}}\bar{q}\,,\nonumber\\
|\hspace{0.09cm}6\hspace{0.09cm}\rangle:&sccq_{[4]_{X}[31]^1_{F}[31]_{S}[211]_{C}}\bar{q}\,,\label{3half}\\
|\hspace{0.09cm}7\hspace{0.09cm}\rangle:&sccq_{[4]_{X}[31]^2_{F}[31]_{S}[211]_{C}}\bar{q}\,,\nonumber\\
|\hspace{0.09cm}8\hspace{0.09cm}\rangle:&sccq_{[4]_{X}[4]_{F}[31]_{S}[211]_{C}}\bar{q}\,,\nonumber
\end{eqnarray}
and for $J^P=3/2^{-}$ or $5/2^{-}$ are
\begin{eqnarray}
|\hspace{0.09cm}9\hspace{0.09cm}\rangle:&~sccq_{[4]_{X}[31]^1_{F}[4]_{S}[211]_{C}}\bar{q}\,,\nonumber\\
|10\rangle:&~sccq_{[4]_{X}[31]^2_{F}[4]_{S}[211]_{C}}\bar{q}\,,
\label{5half}
\end{eqnarray}
respectively.

One should note that the different spin symmetries of the four-quark
subsystem result in vanishing coupling between different five-quark configurations,
this is another reason for us to categorize the states in three groups by
the four-quark spin wave functions.

\subsection{The chiral Lagrangian approach} \label{cla}

\begin{figure*}[htbp]
\begin{center}
\includegraphics[scale=0.65]{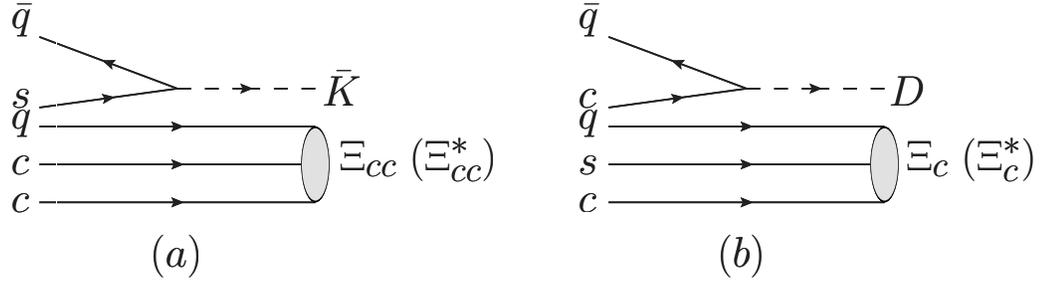}
\caption{Transitions of the $sccq\bar{q}$ states to
$\bar{K}\Xi_{cc}$~($\Xi_{cc}^{*}$) (a) and
$D\Xi_{c}$~($\Xi_{c}^{*}$) (b).} \label{fey}
\end{center}
\end{figure*}

We consider the decays $sccq\bar{q}\rightarrow MB$, which mainly
proceed through the process of $q\bar{q}\rightarrow M$, where the
final baryon and meson are assumed to be composed of three-quark and
a quark-antiquark pair, respectively. We name this kind of decays as
the annihilation transitions. The $sccq\bar{q}$ to $\bar{K}\Xi_{cc}$
and $D\Xi_{cc}$ transitions are shown in Fig.~\ref{fey}.

To compute the transitions of the $\Omega_{cc}\rightarrow\bar{K}\Xi_{cc}$
and $\Omega_{cc}\rightarrow D\Xi_{c}$ shown in
Fig.~\ref{fey}, we use the chiral Lagrangian approach. Within this
approach, the quark pseudoscalar (P) and vector (V) meson couplings are
\begin{eqnarray}
H_{eff}^{Pqq}&=&\sum_{j}\bar{\psi}_{j}\gamma_{\mu}^{j}\gamma_{5}^{j}\psi_{j}\partial^{\mu}\phi_{m}\,,
\label{pqq}\\
H_{eff}^{Vqq}&=&-\sum_{j}\bar{\psi}_{j}(a\gamma_{\mu}^{j}+\frac{ib\sigma_{\mu\nu}k_{M}^{\nu}}{2m_{j}})\phi_{m}^{\mu}\psi_{j}\,,
\label{vqq}
\end{eqnarray}
respectively, where the summation on $j$ runs over the quark in the initial hadron. $\psi_{j}$ represents the quark field, and
$\phi_{m}$ and $\phi_{m}^{\mu}$ are the pseudoscalar and vector meson fields. $m_{j}$
is the constituent mass of the $jth$ quark, while $k_{M}^{\nu}$ denotes the four-momentum of the vector meson. $a$ and $b$ are the vector and tensor coupling constants, respectively.

In the non-relativistic approximation, Eqs.~(\ref{pqq}) and
(\ref{vqq}) lead to the operators for process involving
$q\rightarrow q^{\prime}M$ transitions as following
\begin{widetext}
\begin{eqnarray}
T_{d}^{Pqq}&=&\sum_{j}\left(
\frac{\omega_{M}}{E_{f}+M_{f}}\sigma_j\cdot\vec{P}_{f}+
\frac{\omega_{M}}{E_{i}+M_{i}}\sigma_j\cdot\vec{P}_{i}
-\sigma_j\cdot\vec{k}_{M}
+\frac{\omega_{M}}{2\mu_{q}}\sigma_j\cdot\vec{p}_{j}\right)X^{j}_{P} \exp\{-i\vec{k}_{M}\cdot\vec{r}_{j}\}\,,\label{pseudo}\\
T^{Vqq}_{d,T}&=&\sum_{j}\left\{i\frac{b^{\prime}}{2m_{j}}\vec{\sigma}_{j}\cdot(\vec{k}_{M}\times\vec{\epsilon})
+\frac{a}{2\mu_{q}}\vec{p}_{j}\cdot\vec{\epsilon}\right\}X_{V}^{j}
\exp\{-i\vec{k}_{M}\cdot\vec{r}_{j}\}\,,\label{trans}\\
T^{Vqq}_{d,L}&=&\sum_{j}\frac{aM_{V}}{|\vec{k}_{M}|}X_{V}^{j}\exp\{-i\vec{k}_{M}\cdot\vec{r}_{j}\}\,,\label{long}
\end{eqnarray}
\end{widetext}
where $\omega_{M}$ and $\vec{k}_{M}$ are the energy and
three-momentum of the final meson. $E_{i(f)}$, $M_{i(f)}$ and
$\vec{P}_{i(f)}$ are the energy, mass and three-momentum of the
initial (final) baryon, while $\vec{p}_{j}$, $\vec{r}_{j}$ and
$m_{j}$ are the momentum, coordinate and constituent mass of the
quark which emits a meson. The $\mu_{q}$ is the reduced mass of the
$j$th quark before and after emitting the meson. For the vector
meson emission, in Eqs.~(\ref{trans}) and~(\ref{long}), the
transition operators are denoted as $T^{Vqq}_{d,T}$ and
$T^{Vqq}_{d,L}$ for the meson being transversely and longitudinally
polarized, respectively. The $b^{\prime}$ in Eq.~(\ref{trans}) is
defined by $b^{\prime}=b-a$. $M_{V}$ is the mass of the vector
meson, and the polarization vectors of the final vector meson are
taken to be
\begin{eqnarray}
\epsilon_{\mu}^{L}=\frac{1}{M_{V}}\left(
\begin{array}{c}
 |\vec{k}_{M}| \\
E_{V}\frac{\vec{k}_{M}}{|\vec{k}_{M}|}
\end{array}
\right)\,,~~~~~~~
\epsilon_{\mu}^{T}=\left(
\begin{array}{c}
 0 \\
         \vec{\epsilon}
\end{array}
\right)\,,
\end{eqnarray}
with
\begin{eqnarray}
\vec{\epsilon}(\pm)=1/\sqrt{2}(\mp1,-i,0)^{T}\,,
\end{eqnarray}
and $E_{V}$ is the energy of the final vector meson.

Finally, $X^{j}_{P}$ and $X^{j}_{V}$ are the operators in flavor
space for a pseudoscalar and vector meson emission, which only
depends on the quark-antiquark content of the emitted meson. For
instance, $X^{j}_{P}$ for a light pseudoscalar meson emission in
Eq.~(\ref{pseudo}) can be defined as
\begin{eqnarray}
X^{j}_{\pi^{\pm}} &=& \mp\frac{1}{\sqrt{2}}(\lambda_{1}^{j}\mp
i\lambda_{2}^{j}), \\
X^{j}_{\pi^{0}} &=& \lambda_{3}^{j}\,,\\
X^{j}_{K^{\pm}} &=& \mp\frac{1}{\sqrt{2}}(\lambda_{4}^{j}\mp i\lambda_{5}^{j})\,,\\
X^{j}_{K^{0},\bar{K}^{0}} &=&
\mp\frac{1}{\sqrt{2}}(\lambda_{6}^{j}\mp i\lambda_{7}^{j})\,,\\
X^{j}_{\eta} &=& cos\theta\lambda_{8}^{j}-sin\theta\sqrt{\frac{2}{3}}\mathcal{I}\,,\\
X^{j}_{\eta^\prime} &=&
sin\theta\lambda_{8}^{j}+cos\theta\sqrt{\frac{2}{3}}\mathcal{I}\,,
\end{eqnarray}
with $\lambda_{i}^{j}$ and $\mathcal{I}$ the Gell-Mann matrix and
unit matrix in flavor space. $\theta$ denotes
the mixing angle for the mixing between $\eta_{1}$ and $\eta_{8}$,
leading to the physical states $\eta$ and $\eta^\prime$
\begin{eqnarray}
  \eta &=& \eta_{8}cos\theta-\eta_{1}sin\theta\nonumber \\
  \eta^{\prime} &=& \eta_{8}sin\theta+\eta_{1}cos\theta\,,
\end{eqnarray}
where the empirical value for the mixing angle is $\theta=-23$\textdegree.
The flavor operators for other
pseudoscalar mesons or the vector mesons can be obtained
straightforward.

Accordingly, the transition operators for a pseudoscalar meson
emission $T_{a}^{Pqq}$, a transversely polarized vector meson
emission $T^{Vqq}_{a,T}$ and a longitudinally polarized vector meson
emission can be obtained as
\begin{widetext}
\begin{eqnarray}
T_{a}^{Pqq} &=& \sum_{j}(m_{j}+m_{\bar{q}})
\mathcal{C}^{j}_{XFSC}\bar{\chi}_{z}^{\dag}\mathcal{I}_{2}\chi_{z}^{j}X_{P}^{j}\exp\{-i\vec{k}_{M}\cdot(\vec{r}_{j}+\vec{r}_{\bar{q}})/2\}\,,
\label{tp}\\
T^{Vqq}_{a,T}&=&\sum_{j}\left\{a-\frac{m_{j}+m_{\bar{q}}}{2m_{j}}b\right\}\vec{\sigma}\cdot\vec{\epsilon}X_{V}^{j}\exp\{-i\vec{k}_{M}\cdot(\vec{r}_{j}+\vec{r}_{\bar{q}})/2\}\,,\label{tvt}\\
T^{Vqq}_{a,L}&=&\sum_{j}\left\{a-\frac{m_{j}+m_{\bar{q}}}{2m_{j}}b\right\}\frac{E_{V}\vec{\sigma}\cdot\vec{k}_{M}}{M_{V}|\vec{k}_{M}|}X_{V}^{j}\exp\{-i\vec{k}_{M}\cdot(\vec{r}_{j}+\vec{r}_{\bar{q}})/2\}\label{tvl}\,,
\end{eqnarray}
\end{widetext}
where $m_{j}$ and $m_{\bar{q}}$ are the constituent masses of the
$jth$ quark and the antiquark, respectively.
$\mathcal{C}^{j}_{XFSC}$ denotes the operator to calculate the
orbital, flavor, spin and color overlap factor between the residual
wave function of the pentaquark configuration after the
quark-antiquark annihilation and the wave function of the final
baryon. $\bar{\chi}_{z}^{\dag}\mathcal{I}_{2}\chi_{z}^{j}$ is the
spin operator for the quark-antiquark annihilation.

%

\section{Numerical results and discussions}
\label{num}

In this section, we present our theoretical results for the mass
spectrum of the low-lying $sccq\bar{q}$ states with $J^{P}=1/2^{-}$,
$3/2^{-}$ and $5/2^{-}$ , and the decay
behaviours of the obtained $\Omega_{cc}$ pentaquark states.

\subsection{The mass spectrum of the low-lying $sccq\bar{q}$ states}
\label{rspec}

With the Hamiltonian in Eq.~(\ref{ham5}) and wave function in
Eq.~(\ref{wfc}), one can obtain the following nonzero
$H_{ij}=\langle i|H|j\rangle$ matrix elements
\begin{eqnarray}
H_{11}&=&E_{0}-7.50C_{D}-7.50C_{D_{s}}\,,\nonumber\\
H_{12}&=&H_{21}=4.90C_{D}-4.90C_{D_{s}}\,,\nonumber\\
H_{13}&=&H_{31}=-4.33C_{K}+0.87C_{D}+0.87C_{D_{s}}+2.60C_{c\bar{c}}\,,\nonumber\\
H_{22}&=&E_{0}-1.50C_{K}-2.00C_{D}-2.00C_{D_{s}}-1.50C_{c\bar{c}}\,,\nonumber\\
H_{23}&=&H_{32}=-5.66C_{D}+5.66C_{D_{s}}\,,\nonumber\\
H_{33}&=&E_{0}-5.0C_{K}-2.50C_{D}-2.50C_{D_{s}}+3.0C_{c\bar{c}}\,,
\label{enq1}
\end{eqnarray}
between the pentaquark configurations in Eq.~(\ref{1half}), and
\begin{eqnarray}
H_{44}&=&E_{0}-2.5C_{K}-4.50C_{D}-4.50C_{D_{s}}-1.5C_{c\bar{c}}\,,\nonumber\\
H_{45}&=&H_{54}=-6.12C_{D}+6.12C_{D_{s}}\,,\nonumber\\
H_{46}&=&H_{64}=-3.54C_{D}+3.54C_{D_{s}}\,,\nonumber\\
H_{47}&=&H_{74}=5.00C_{K}-2.50C_{D}-2.50C_{D_{s}}\,,\nonumber\\
H_{55}&=&E_{0}-0.50C_{K}-4.00C_{D}-4.00C_{D_{s}}-0.5C_{c\bar{c}}\,,\nonumber\\
H_{56}&=&H_{65}=1.73C_{K}-1.73C_{c\bar{c}}\,,\nonumber\\
H_{57}&=&H_{75}=-1.22C_{D}+1.22C_{D_{s}}\,,\nonumber\\
H_{58}&=&H_{85}=-1.41C_{K}+1.41C_{D}+1.41C_{D_{s}}-1.41C_{c\bar{c}}\,,\nonumber\\
H_{66}&=&E_{0}+1.50C_{K}-4.0C_{D}-4.0C_{D_{s}}+1.50C_{c\bar{c}}\,,\nonumber\\
H_{67}&=&H_{76}=-0.71C_{D}+0.71C_{D_{s}}\,,\nonumber\\
H_{68}&=&H_{86}=-2.45C_{K}+2.45C_{c\bar{c}}\,,\nonumber\\
H_{77}&=&E_{0}-2.5C_{K}-0.50C_{D}-0.50C_{D_{s}}-1.5C_{c\bar{c}}\,,\nonumber\\
H_{78}&=&H_{87}=-3.46C_{D}+3.46C_{D_{s}}\,,\nonumber\\
H_{88}&=&E_{0}+0.50C_{K}+1.0C_{D}+1.0C_{D_{s}}+0.50C_{c\bar{c}}\,,
\label{enq2}
\end{eqnarray}
between the pentaquark configurations in Eq.~(\ref{3half}), and
\begin{eqnarray}
H_{99}&=&E_{0}-1.50C_{K}+1.0C_{D}+1.0C_{D_{s}}-1.50C_{c\bar{c}}\,,\nonumber\\
H_{910}&=&H_{109}=2.83C_{D}-2.83C_{D_{s}}\,,\nonumber\\
H_{1010}&=&E_{0}+2.50C_{K}-C_{D}-C_{D_{s}}-1.50C_{c\bar{c}}\,,
\label{enq3}
\end{eqnarray}
between the pentaquark configurations in Eq.~(\ref{5half}).

In the above equations, $C_{M}$
are the corresponding matrix elements of the hyperfine interaction
coupling strength $V_{M}(r_{ij})$ between the S-wave orbital wave
functions of the quarks in $sccq\bar{q}$ system, namely
\begin{equation}
C_{M}=\langle\varphi(\{\vec{\xi}_{j}\})|V_{M}(r_{ij})|\varphi(\{\vec{\xi}_{j}\})\rangle\,.
\end{equation}
$C_{c\bar{c}}$ is obtained from the last term in
Eq.~(\ref{hyp}), and it contains the exchanges of the $u\bar{u}$,
$d\bar{d}$, $s\bar{s}$ and $c\bar{c}$ pairs. The coupling strength
constants $C_{M}$ are taken to be the empirical
values~\cite{Glozman:1995fu} as shown in Table~\ref{cm}.
\begin{table}[htbp]
\caption{\footnotesize The hyperfine interaction coupling strength constants (in MeV).
\label{cm}}
\renewcommand
\tabcolsep{0.30cm}
\renewcommand{\arraystretch}{1.2}
\begin{tabular}{cccccc}
\toprule[1.2pt]
$C_{\pi}$ & $C_{K}$ & $C_{s\bar{s}}$ & $C_{D}$ &  $C_{D_{s}}$ & $C_{c\bar{c}}$ \\

\hline

$21.0$ & $15.5$ & $11.5$ & $6.5$ &  $6.5$ & 0 \\
\bottomrule[1.2pt]
\end{tabular}
\end{table}
%
%
\begin{figure}[htbp]
\begin{center}

\includegraphics[scale=0.45]{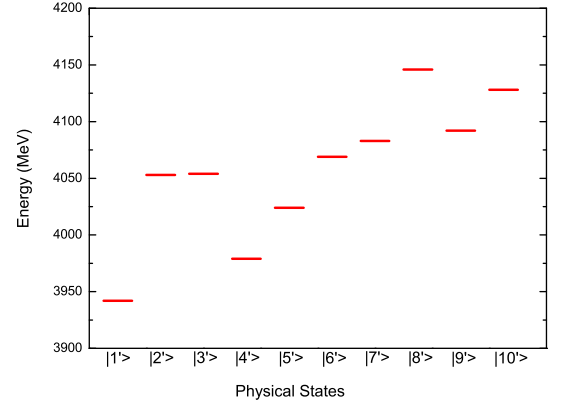}
\end{center}
\vspace{-3.5cm} \caption{\footnotesize Spectrum of the obtained
physical states.} \label{spec}
\end{figure}

%
\begin{table*}[htbp]
\caption{\footnotesize The ten physical pentaquark states obtained
in present model, line three shows the energies for the states
$|i^{\prime}\rangle$ (in MeV), and lines four to ten are the
corresponding probability amplitudes. \label{nmm}}
\renewcommand
\tabcolsep{0.30cm}
\renewcommand{\arraystretch}{1.5}
\begin{tabular}{c|ccc|ccccc|cc}

\toprule[1.2pt]

& \multicolumn{3}{c}{$J^{P}=1/2^{-}$} & \multicolumn{5}{c}{$J^{P}=1/2^{-}$~or~$3/2^{-}$}  & \multicolumn{2}{c}{$J^{P}=3/2^{-}$~or~$5/2^{-}$}\\

\hline

& $|1^{\prime}\rangle$ & $|2^{\prime}\rangle$ & $|3^{\prime}\rangle$
& $|4^{\prime}\rangle$ &   $|5^{\prime}\rangle$ &
$|6^{\prime}\rangle$
&  $|7^{\prime}\rangle$ &  $|8^{\prime}\rangle$ &  $|9^{\prime}\rangle$ &  $|10^{\prime}\rangle$ \\

            & 3942    &  4053    &   4054  &  3979    & 4024   &  4069  & 4083   & 4146   & 4092    & 4128    \\
\hline

$|1\rangle$ &  0.67   &   0      &   0.75   &  0      &  0     &   0    &  0     &  0     &  0      &   0    \\

$|2\rangle$ &    0    &     1    &   0      &   0     & 0      &   0    &  0     &  0     &  0      &   0    \\

$|3\rangle$ &   0.75  &     0    &   -0.67  &  0      &  0     &  0     &  0     &  0     &  0      &   0    \\

$|4\rangle$ &    0    &      0   &    0     &  0.87   &   0    &   0    &  -0.50 &  0     &   0     &   0    \\

$|5\rangle$ &    0    &     0    &     0    &   0     &   0.79 & 0.59   &  0     &  -0.16 &   0     &   0     \\

$|6\rangle$ &   0     &     0    &    0     &   0     & -0.58   & 0.64  &  0     & -0.50  &   0     &   0    \\

$|7\rangle$ &   0     &     0    &     0    &  -0.50  &   0     & 0      & -0.87 &  0     &   0     &   0    \\

$|8\rangle$ &   0     &    0     &    0     &  0      &  -0.19  & 0.49   &  0    &  0.85  &   0     &   0    \\

$|9\rangle$ &    0    &    0     &    0     &    0    &   0     &   0    &  0    &  0     &   1     &   0    \\

$|10\rangle$ &    0   &   0      &   0      &   0     &   0     &  0     &   0   &  0     &   0     &   1    \\

\bottomrule[1.2pt]

\end{tabular}
\end{table*}

With the above values for the model parameters and the
diagonalization of the matrices obtained by
Eqs.~(\ref{enq1}--\ref{enq3}), one can get the physical states which
are shown in Fig.~\ref{spec}, while the explicit probability
amplitudes are shown in Table~\ref{nmm}. For instances,
Eq.~(\ref{enq1}) leads to the following energy matrix
\begin{equation}
E=\left(
\begin{array}{ccc}
4005& 0 & -55.8 \\
  0 & 4053 & 0   \\
 -55.8 & 0 & 3992 \\
\end{array}
\right)\,\mathrm{Mev}\,.\label{A3}
\end{equation}
Then one can directly obtain the eigenvalues and eigenvectors of matrix in Eq.~(\ref{A3}).
The three obtained eigenvalues are the energies of the physical
states $|i^\prime\rangle$ with $i=1,2,3$, respectively, and a obtained
eigenvector just show the coefficients for the decoupling of a corresponding
physical state $|i^\prime\rangle$ to the configurations $|i\rangle$ listed in Eq.~(\ref{1half}).

 The energies for the
obtained $\Omega_{cc}$ states in present work are at
$4050 \pm 100$~MeV. Similar as the results for $\Omega_{c}^{0}$
obtained in Ref.~\cite{An:2017lwg}, mixing between the pentaquark
configurations $|i\rangle$ caused by the goldstone boson exchange is
strong, while the mass splitting for the obtained states
$|i^{\prime}\rangle$ is not very large. On the other hand, the
spectrum of the ten obtained states is not sensitive to the values
of the coupling strength $C_{M}$.

%



Up to now, there are no solid experimental data for the
$\Omega_{cc}$ resonances, while theoretical investigations on the
doubly heavy baryon resonances have been intensively taken using
various of approaches, such as the constituent quark
model~\cite{Lu:2017meb,Xiao:2017udy,Ebert:2002ig,Albertus:2006ya,Faessler:2009xn,Karliner:2014gca,Li:2019tbn},
QCD sum rules~\cite{Chen:2017sbg}, chiral perturbation
theory~\cite{Guo:2017vcf,Yao:2018ifh,Li:2017pxa}, the unitarized
coupled channel approach~\cite{Yan:2018zdt}, and the lattice QCD
calculations~\cite{Mathur:2018rwu}, etc. The corresponding obtained
energies for the $P$-wave $\Omega_{cc}$ in a three-quark picture are
around $4000-4200$~MeV in most of the literatures, and one may note
that in Ref.~\cite{Yan:2018zdt}, the $S$-wave interactions between
pseudo-Nambu-Goldstone bosons ($\pi$, $K$ and $\eta$) and the
$J^{P}=1/2^{+}$ ground state doubly charmed baryons in the energy
region around the corresponding thresholds are investigated, two
quasistable narrow $J^{P}=1/2^{-}$ $\Omega_{cc}$ are predicted to
lie at the energy below $4200$~MeV, and their strong decay mode is
predicted to be only the $\Omega_{cc} \pi^{0}$, which is isospin
breaking channel. Therefore, the two obtained $\Omega_{cc}$ resonances
in a meson-baryon picture should be very narrow.

One may also study the spectrum of low-lying
$\Omega_{cc}$ resonances with negative parity using the chiral
constituent quark model in a three-quark
picture~\cite{Charmhyperons}. As $P$-wave states whose parity are
negative, there are three possible $\Omega_{cc}$ configurations:
\begin{eqnarray}
|scc,1\rangle&=&[21]_{X}[21]_{FS}[21]_{F}[21]_{S}[1^3]_C\,,\nonumber\\
|scc,2\rangle&=&[21]_{X}[21]_{FS}[3]_{F}[21]_{S}[1^3]_C\,,\label{3qconfig}\\
|scc,3\rangle&=&[21]_{X}[21]_{FS}[21]_{F}[3]_{S}[1^3]_C\,,\nonumber
\end{eqnarray}
whose spin-parity quantum number $J^P$ may be $1/2^-$ or $3/2^-$ for
the first two configurations, and $1/2^-$, $3/2^-$ or $5/2^-$ for
the last one. Direct calculations employing the chiral constituent
quark model as in Ref.~\cite{Charmhyperons} lead to the following
values for the energies of the three $\Omega_{cc}$ states,
\begin{equation}
E_1 =
4219\,\mathrm{MeV},\hspace{0.15cm}E_2=4246\,\mathrm{MeV},\hspace{0.15cm}E_3=4257\,\mathrm{MeV},\label{q3num}
\nonumber
\end{equation}
respectively. Consequently, the energies of low-lying $\Omega_{cc}$
states in the five-quark picture are lower than those in the
three-quark picture, this conclusion is the same as that for the
$\Omega^{*}$ resonances~\cite{Yuan:2012zs}.

In Ref.~\cite{Ebert:2002ig}, a relativistic quark model was applied
to study the spectrum of doubly heavy baryons. Considering the $\Omega_{cc}$
resonances to be dominated by three-quark components, it was obtained that the
low-lying $\Omega_{cc}$ resonances with negative parity fall in the range of $4200-4300$~MeV,
which are consistent with the results obtained in Ref.~\cite{Xiao:2017udy}
by employing a three-quark model.

While in Ref.~\cite{Lu:2017meb}, a three-quark model was employed
to investigate the spectrum of the doubly heavy baryons,
in which model the two heavy quarks were treated as a diquark, and the
resulting energies of the low-lying $\Omega_{cc}$ were in the range of $4050-4150$~MeV.
Those results are about $100$~MeV lower than the present rough estimation using
a three-quark model, and the results in~\cite{Xiao:2017udy,Ebert:2002ig}.
So one may expect that the diquark assumption for the two heavy quark
in $\Omega_{cc}$ resonances may reduce the energies.

In any case, we can conclude that the $\Omega_{cc}$ resonances should lie at
a energy below $4200$~MeV in both the compact five-quark model (present) and
the meson-baryon model~\cite{Yan:2018zdt}.

Finally, we show the dependency of presently obtained spectrum on the model parameter
$E_{0}$. By taking $E_{0}=3132$~MeV as given in Ref.~\cite{An:2017lwg}, one can get
\begin{equation}
E_{i^{\prime}}\simeq 3075\pm100\,\mathrm{MeV}\,.
\end{equation}
Obviously, the obtained energies are much lower than those predicted by using other approaches.
Namely, the value $E_{0}=4102$~MeV employed in our calculations should be reasonable. We
also present the numerical results with $E_{0}$ changed by $2\%$, then
\begin{equation}
E_{i^{\prime}}\simeq 4125\pm100\,\mathrm{MeV}\,.
\end{equation}
In fact, change of $E_{0}$ should lead to almost the same change for energy of each physical state $|i^{\prime}\rangle$.
In addition, the coefficients for decompositions of the physical states $|i^{'}\rangle$
are not sensitive to $E_{0}$.

\subsection{S-wave coupling of the $sccq\bar{q}$ to pseudoscalar meson and ground baryon states}
\label{rcou}

From Fig.~\ref{spec} and Table~\ref{nmm}, one can find that most of the
obtained physical $sccq\bar{q}$ states are above the threshold of
the $SU(2)$ isospin breaking $\pi\Omega_{cc}$ channel, but, below the
thresholds of the other pseudoscalar meson and ground state baryons
channels. It is expected that the decay widths of the presently
obtained $\Omega_{cc}$ should not be very large. This is in
consistent with these findings in Ref.~\cite{Yan:2018zdt}.
Therefore, in present work, we only try to estimate the $S$-wave
transitions of the obtained $1/2^{-}$ and $3/2^{-}$ $\Omega_{cc}$
resonances to $MB$ channels.

Using the transition operator in Eq.~(\ref{tp}), and the wave functions obtained in Sec.~\ref{rspec},
one can calculate the transition matrix elements of the obtained
$\Omega_{cc}$ resonances to $\bar{K}\Xi_{cc}$,
$\bar{K}\Xi_{cc}^{*}$, $D\Xi_{c}$, and $D\Xi_{c}^{*}$
channels, respectively. It is found that all the transition
amplitudes of $\Omega_{cc} \to MB$ processes share a common factor
which involves the overlap between the orbital wave functions of the
pentaquark configurations and the final meson-baryon, namely,
\begin{widetext}
\begin{equation}
\mathcal{F}(k^{2}_{M})=\langle\phi_{B}(\{\vec{\xi}_{i}^{\prime}\})|\exp
\{-i\vec{k}_{M} \cdot
(\vec{r}_{j}+\vec{r}_{\bar{q}})/2\}|\varphi(\{\vec{\xi}_{i}^{\prime}\})\rangle\,,
\end{equation}
\end{widetext}
which depends on the momentum of the final meson $\vec{k}_{M}$, and
the explicit confinement potential model for the quarks in baryons.
Note that in the transitions of almost all the
presently obtained $\Omega_{cc}$ resonances to the corresponding $MB$ channels, the
$\Omega_{cc}$ resonances are off-shell, because of their lower
energies than the mass thresholds of the $MB$ channels. Thus, the
momentum of the final meson $\vec{k}_{M}$ in Eq.~(\ref{tp}) cannot
be pinned down in present framework. Yet, one can estimate the
partial decay widths of these obtained $\Omega_{cc}$ resonance from
the calculation of the flavor, spin, orbital, and color overlap
factor for the final $MB$ states and the residual three-quark-meson
configurations of the $\Omega_{cc}$ states after the annihilation of
the quark-antiquark $q\bar{q}\rightarrow M$, namely, the transition
matrix elements of the orbital-flavor-spin-color dependent operator
$\mathcal{C}^{j}_{XFSC}\bar{\chi}_{z}^{\dag}\mathcal{I}_{2}\chi_{z}^{j}X_{P}^{j}$
in Eq.~(\ref{tp}).

In the three quark model, for $\Xi_{c}$ baryon,
the light and strange quarks in its
flavor wave function can be either symmetric or antisymmetric.
We denote the former sate as $\Xi_{c}$, while the
latter one as $\Xi_{c}^{\prime}$. Therefore, we consider the
transition processes of $\Omega_{cc}$ with spin-parity $1/2^{-}$ to
$\bar{K}\Xi_{cc}$, $D\Xi_{c}$, and $D\Xi_{c}^{\prime}$
channels, and the $\Omega_{cc}^{*}$ with spin-parity $3/2^{-}$ to
$\bar{K}\Xi_{cc}^{*}$ and $D\Xi_{c}^{*}$ channels.

%
\begin{table*}[ht]
\caption{\footnotesize Color-flavor-spin factors for the transitions
$J^{P}=1/2^{-}$ (shown in the $2^{nd}$ to $4^{th}$ rows) and
$3/2^{-}$ (shown in the last two rows) $sccq\bar{q}$ configurations
$|i\rangle$ to $MB$ channels. \label{OFC}}
\renewcommand
\tabcolsep{0.30cm}
\renewcommand{\arraystretch}{1.5}
\begin{tabular}{c|cccccccccc}
\toprule[1.2pt]

            & $|1\rangle$   & $|2\rangle$   & $|3\rangle$   & $|4\rangle$   &   $|5\rangle$ &  $|6\rangle$  &  $|7\rangle$  &  $|8\rangle$ & $|9\rangle$   &  $|10\rangle$\\
\hline

$\bar{K}\Xi_{cc}$ &  $-1$         & $\sqrt{6}/3$  &  $\sqrt{3}/3$ &  $\sqrt{3}$   &  $\sqrt{2}$   &  $\sqrt{6}/3$ &  $\sqrt{3}/3$ &  $0$  &  $-$ &               $-$  \\

$D\Xi_{c}$  & $0$           & $\sqrt{12}/3$ &  $0$          &  $0$          &  $-2$         & $\sqrt{12}/3$ &  $0$          &  $0$   &  $-$ &                $-$  \\

$D\Xi_{c}'$& $\sqrt{6}/3$  & $0$           &  $\sqrt{2}$   &  $-\sqrt{2}$  &  $0$          & $0$           &  $\sqrt{2}$   &  $0$  & $-$   &                 $-$  \\

\hline

$\bar{K}\Xi_{cc}^{*}$ &   $-$       & $-$  &  $-$  &  $0$& $0$        & $-\sqrt{6}/3$ & $\sqrt{12}/3$ &  $1$        & $\sqrt{15}/3$  & $-\sqrt{30}/3$   \\

$D\Xi_{c}^{*}$  &   $-$      & $-$ & $-$  & $0$  & $0$        & $\sqrt{12}/3$ &     $0$       & $\sqrt{2}$  & $-\sqrt{30}/3$ & $0$   \\

\bottomrule[1.2pt]
\end{tabular}
\end{table*}
%

Then, straightforward calculations on the transition matrix elements
for the operator
$\mathcal{C}^{j}_{XFSC}\bar{\chi}_{z}^{\dag}\mathcal{I}_{2}\chi_{z}^{j}X_{P}^{j}$
in the processes of $|i\rangle$ in Eqs.~(\ref{1half}-\ref{3half}) to
the above mentioned $MB$ channels lead to the results shown in
Table~\ref{OFC}, where the first three rows of the numerical results
are the orbital-flavor-spin-color overlap factors for the
configurations $|i\rangle$ with spin-parity quantum number $1/2^{-}$
to $MB$ channels, while the last two rows are those for $3/2^{-}$
configurations.

Considering the probability amplitudes for the mixing of
configurations $|i\rangle$ presented in Table~\ref{nmm}, we obtain
the corresponding overlap factors for the physical $\Omega_{cc}$
resonances listed in Table~\ref{OFP}. One may note that there are
some zeros obtained for some configurations as shown in Table~\ref{OFC},
however, they become finite for the physical states. For instance, the
overlap factor for the configuration $|8\rangle$ to the channel $\bar{K}\Xi_{cc}$
is $0$, while that for $|8^\prime\rangle\rightarrow\bar{K}\Xi_{cc}$ is $-0.635$.
This is because of that the physical state $|8^\prime\rangle$ decouples
to the configurations $|i\rangle$ as
\begin{equation}
|8^\prime\rangle=-0.16|5\rangle-0.50|6\rangle+0.85|8\rangle\,,
\end{equation}
which is shown in Table~\ref{nmm}.

Compared to the overlap
factor~\cite{An:2018vmk,An:2009uv} for the strangeness five-quark
configurations $|uuds\bar{s}\rangle$ to $\eta p$
channels that is about $\sim0.75$, which may account for the strong coupling between
$S_{11}(1535)$ and strangeness channels, one can expect that the
presently obtained physical states $|i^{\prime}\rangle$ may couple strongly to the
$MB$ channels for which the overlap factors shown in Table~\ref{OFP} are larger than $0.8$.

It should be very interesting to compare the decay behaviours of the presently obtained $\Omega_{cc}$ resonances
with those in a three-quark model. In Ref.~\cite{Lu:2017meb}, five $\Omega_{cc}$ resonances
lying at $4208-4303$~MeV were obtained, and the decay widths of these resonances to $\bar{K}\Xi_{cc}$
or $\bar{K}\Xi_{cc}^{*}$ channels were estimated explicitly. It was found that some of the
obtained decay widths should be larger than $100$~MeV. This is very different from the conclusion
that most of the obtained $\Omega_{cc}$ resonances using a pentaquark picture
can only decay to the isospin breaking channel $\pi\Omega_{cc}$.

In a three-quark picture, one can also estimate the flavor-spin-color overlap factor of the
$\Omega_{cc}$ resonances and the $MB$ channels using Eq.~(\ref{pseudo}).
For instance, a straightforward calculation on the overlap factors of the three-quark states
given in Eq.~(\ref{3qconfig}), shows that coupling for $\Omega_{cc}\rightarrow\bar{K}\Xi_{cc}$
may be comparable to that for $\Omega_{cc}\rightarrow D\Xi_{c}$, since the obtained
flavor-spin-color overlap
factor for a given three-quark $\Omega_{cc}$ resonance to the $\bar{K}\Xi_{cc}$ channel
is $\sqrt{2}$ times of that for the $D\Xi_{c}$ channel, this is determined by
the flavor-spin structure of the $\Omega_{cc}$ resonances and the effective chiral
Lagrangian. However, as we can see in Table~\ref{OFP}, the presently obtained numerical results
for several states are very different from the three-quark results.

\begin{table*}[ht]
    \caption{\footnotesize Color-flavor-spin overlap factors for the transitions $J^{P}=1/2^{-}$
(shown in the $2^{nd}$ to $4^{th}$ rows) and $3/2^{-}$ (shown in the
last two rows) $sccq\bar{q}$ physical states $|i'\rangle$ to $MB$
channels.
        \label{OFP}}
    \renewcommand
    \tabcolsep{0.30cm}
    \renewcommand{\arraystretch}{1.5}
    \begin{tabular}{c|cccccccccc}
\toprule[1.2pt]

        & $|1'\rangle$   & $|2'\rangle$   & $|3'\rangle$   & $|4'\rangle$   &   $|5'\rangle$ &  $|6'\rangle$  &  $|7'\rangle$  &  $|8'\rangle$ & $|9'\rangle$ & $|10'\rangle$ \\
        \hline

        $\bar{K}\Xi_{cc}$ &  $-0.237$         & $0.816$  &  $-1.137$ &  $1.218$   &  $0.644$   &  $1.357$ &  $-1.368$ &  $-0.635$                 & $-$  & $-$ \\

        $D\Xi_{c}$  & $0$           & $1.155$ &  $0$          &  $0$          &  $-2.250$         & $-0.441$ &  $0$          &  $-0.257$                   & $-$  &  $-$ \\

        $D\Xi_{c}'$& $1.608$  & $0$           &  $-0.335$   &  $-1.937$  &  $0$          & $0$           &  $-0.523$   &  $0$                   & $-$  & $-$ \\

\hline

$\bar{K}\Xi_{cc}^{*}$ & $-$ & $-$ & $-$ & $-0.577$        &  $0.284$        & $-0.033$ & $-1.005$ &  $1.258$        & $1.291$  & $-1.826$   \\

        $D\Xi_{c}^{*}$ & $-$ & $-$ & $-$  &  $0$        &  $-0.938$        & $1.432$ &     $0$       & $0.625$  & $-1.826$ & $0$   \\

\bottomrule[1.2pt]
    \end{tabular}
\end{table*}

\section{Summary}
\label{sumcon}
In present work, we investigate the spectrum of low-lying
$\Omega_{cc}$ resonances with negative parity as pentaquark states,
using the chiral constituent quark model within a five-quark
picture. We obtain ten pentaquark states with spin-parity
$J^{P}=1/2^{-},~3/2^{-},~5/2^{-}$, which lie at $4050\pm 100$~MeV.
Most of the obtained states are above the isospin breaking decay
channel $\pi \Omega_{cc}$, but below the other meson-baryon
channels. So we just try to calculate the flavor, spin, orbital, and
color overlap factor for the final $MB$ states and the residual
three-quark-meson configurations of the $\Omega_{cc}$ states after
the annihilation of the quark-antiquark $q\bar{q} \rightarrow M$.
It is found that several ones of the presently obtained $\Omega_{cc}$
may couple strongly to $D\Xi_{c}$ or $\bar{K}\Xi_{cc}$ channels.
One may expect that
these calculations here could be compared with the future
experimental measurements which are likely to be done by Belle II
and/or LHCb.

%
\begin{acknowledgments}

We thank Yun-Xia Lang for her contributions at the very beginning of
present work. 
This work is partly supported by the National
Natural Science Foundation of China under Grant Nos. 11675131,
12075288, 12075133, 11735003, 11961141012 and 11835015. It is also
supported by
the Youth Innovation Promotion Association CAS, Taishan
Scholar Project of Shandong Province (Grant No.tsqn202103062),
the Higher Educational Youth Innovation Science and Technology
Program Shandong Province (Grant No. 2020KJJ004),
and the Chongqing Natural Science
Foundation under Project No. cstc2019jcyj-msxmX0409.

\end{acknowledgments}


%
%

\end{document}